\documentclass[reprint,superscriptaddress,amsmath,amssymb,aps,pre]{revtex4-1}

\usepackage[top=2.5cm,bottom=2.5cm,left=1.3cm,right=1.3cm]{geometry}
\usepackage{graphicx}
\usepackage{amssymb,amsmath}
\usepackage{color}

\newcommand{\figref}[1]{Fig.~\ref{fig:#1}}
\renewcommand{\eqref}[1]{Eq.~(\ref{eq:#1})}

\begin{document}

  \title{Minimal cellular automaton model with heterogeneous cell
    sizes \\ predicts epithelial colony growth}

\author{Steffen Lange}
\affiliation{DataMedAssist Group, HTW Dresden-University of Applied
  Sciences, 01069 Dresden, Germany}
\affiliation{OncoRay-National Center for Radiation Research in Oncology, Faculty of Medicine and University Hospital Carl Gustav Carus, Technische Universität Dresden, Helmholtz-Zentrum Dresden-Rossendorf, 01307 Dresden, Germany}

\author{Jannik Schmied}
\affiliation{DataMedAssist Group, HTW Dresden-University of Applied Sciences, 01069 Dresden, Germany}
\affiliation{Faculty of Informatics/Mathematics, HTW Dresden-University of Applied Sciences, 01069 Dresden, Germany}

\author{Paul Willam}
\affiliation{DataMedAssist Group, HTW Dresden-University of Applied Sciences, 01069 Dresden, Germany}

\author{Anja Voss-B\"ohme}
\affiliation{DataMedAssist Group, HTW Dresden-University of Applied Sciences, 01069 Dresden, Germany}
\affiliation{Faculty of Informatics/Mathematics, HTW Dresden-University of Applied Sciences, 01069 Dresden, Germany}

\date{\today}

\begin{abstract}
  Regulation of cell proliferation is a crucial aspect of tissue
  development and homeostasis and plays a major role in morphogenesis,
  wound healing, and tumor invasion. A phenomenon of such regulation
  is contact inhibition, which describes the dramatic slowing of
  proliferation, cell migration and individual cell growth when
  multiple cells are in contact with each other. While many
  physiological, molecular and genetic factors are known, the
  mechanism of contact inhibition is still not fully understood. In
  particular, the relevance of cellular signaling due to interfacial
  contact for contact inhibition is still debated. Cellular automata
  (CA) have been employed in the past as numerically efficient
  mathematical models to study the dynamics of cell ensembles, but
  they are not suitable to explore the origins of contact inhibition
  as such agent-based models assume fixed cell sizes. We develop a
  minimal, data-driven model to simulate the dynamics of planar cell
  cultures by extending a probabilistic CA to incorporate size changes
  of individual cells during growth and cell division. We successfully
  apply this model to previous in-vitro experiments on contact
  inhibition in epithelial tissue: After a systematic calibration of
  the model parameters to measurements of single-cell dynamics, our CA
  model quantitatively reproduces independent measurements of
  emergent, culture-wide features, like colony size, cell density and
  collective cell migration. In particular, the dynamics of the CA
  model also exhibit the transition from a low-density confluent
  regime to a stationary postconfluent regime with a rapid decrease in
  cell size and motion. This implies that the volume exclusion
  principle, a mechanical constraint which is the only inter-cellular
  interaction incorporated in the model, paired with a size-dependent
  proliferation rate is sufficient to generate the observed contact
  inhibition. We discuss how our approach enables the introduction of
  effective bio-mechanical interactions in a CA framework for future
  studies.
\end{abstract}

\maketitle

Understanding the principles of dynamic tissue organization requires
an interdisciplinary effort where quantitative biological observations
must be challenged by mathematical models that integrate existent
insight on a theoretical level~\cite{SchBru2023,
  BruArlFinRonRaeBro2021}. Cell-based mathematical models essentially
consider a tissue as 'a society of cells'~\cite{Vir1855} where cells
behave and interact with each other and the extracellular environment
according to individual cellular characteristics summarizing the
effects of sub-cellular processes, such giving rise to complex
emergent processes at tissue scale. Existing cell-based model
frameworks can be classified into three broad categories,
(i)~continuum descriptions of tissue movement which rarely describe
individual cells explicitly~\cite{AlaHatLowVoi2015, KuaGuaTanZha2023,
  Shr2005, BasRisJoaSasPro2009, HeiAleLaCZajKosCoh2020, KhaStr2021},
(ii)~models that include an explicit description of the cell membrane,
such as the cellular Potts model~\cite{GraGla1992, LiLow2014, Vos2012,
  Kab2012}, vertex models~\cite{NesGodWooJen2018, AltGanSal2017,
  FarRoeAigEatJue2007, KurBroZarBakFle2015,
  MorStuSabLauQueTorChaSuhRam2017, FleOstBakShv2014} or disk
models~\cite{MorStuSabLauQueTorChaSuhRam2017, SchMolYam2020}, and
finally (iii)~particle based-models or cellular automata where cells
are represented as point-like particles~\cite{BoeHatVosCavHerDeu2015,
  Vos2012, TalCavVosDeu2017, RosBoeLanVos2021, RehKliDeuVos2017,
  LanMogHueVos2022, HofVosRinBru2017, FraAlaBoeVosLan2022,
  SepPetCocGraSilHak2013}. Each class has its own merits and
drawbacks: (i)~Tissue scale descriptions allow to capture biophysical
principles which govern large scale phenomena like tissue fluidity and
deformations (elastic, plastic, viscoelastic) or jamming
transitions~\cite{YanBiCzaMerManMar2017, GniSchHal2019}. However,
relating biophysical parameters of cell colonies to measurable single
cell properties remains challenging and simulation of large
collectives with single-cell resolution can cause high computational
cost~\cite{AlaHatLowVoi2015, KuaGuaTanZha2023}. (ii)~Models that
describe cells as spatially extended objects, such as cellular Potts
and vertex models, can incorporate heterogeneous cell sizes and shapes
within the tissue and provide a related contact network for specifying
intercellular interaction. However, the cellular behavior and
interaction is often described indirectly via membrane adjustments
controlled by free energy terms, the latter also involving purely
technical parameters which have an impact on the emergent
dynamics~\cite{KurBakFle2017, Vos2012, BeaKirHenGraBru2023}. This fact
together with the considerable computational effort for full
exploration of the parameter space limits the calibration to concrete
experimental data. (iii)~Cellular automata and interacting particle
systems are computationally more efficient while being mathematically
easier to analyze~\cite{LiePalJagDra2015}. They allow to describe
cellular behavior and intercellular interaction in a rule-based manner
based on clear mechanistic hypotheses and with biologically meaningful
parameters~\cite{OsbFlePitMaiGav2017}. A major drawback, however, is
the idealization of point-like cells hampering the modeling of tissue
dynamics where cell sizes and shapes are supposed to play an essential
role~\cite{AleTre2020}.

Here, we consider monolayers of growing epithelial tissue viewed as a
dynamic 2D arrangement of individual cells, which grow in size, divide
and migrate in response to mechanical constraints and, potentially,
interaction on cell-cell contact. In-vitro experiments on the
development of epithelial tissue of Madin-Darby canine kidney (MDCK)
cells from Puliafito et al. (2012)~\cite{PulHufNevStrSigFygShr2012}
show two distinct growth regimes: At early times, in the confluent
phase, cell density is low and spatially homogeneous and the colony
grows exponentially in time. At longer times, in the postconfluent
phase, cell density increases inside the colony and subexponential
colony growth driven by more motile cells near the edge of the
expanding colony is observed. The inhibition of cellular proliferation
and migration in crowded environments as observed in the center of the
growing monolayers is called contact inhibition of proliferation
response and migration. It was found to be a not only a consequence of
cell-cell contact but of mechanical constraints that cause successive
cell divisions to reduce cell area. To further differentiate between
the impact of biomechanical constraints and that of intercellular
interaction on cell-cell contact, cell-based mathematical modeling was
exploited. Puliafito et al. (2012)~\cite{PulHufNevStrSigFygShr2012}
used a 1D vertex model to support their findings qualitatively.
However, the model parameters and the observables of the emergent
behavior were not systemically related to experimental measurements.
Aland et al. (2015)~\cite{AlaHatLowVoi2015} could reproduce the
observed cell culture topology and radial distribution function with a
hybrid continuum-discrete model, however, the spatial scale of the
culture had to be extrapolated to realistic culture sizes due to
computational cost of the model. Jain et al.
(2022)~\cite{JaiWenVoi2022} incorporated proliferation in the
energetic description of a multiphase field model to predict the
coordination numbers of cells in growing epithelial colonies
corresponding to different hypotheses on contact inhibition and to
confirm the typical, experimentally observed linear boundary growth of
the colony radius. The role of cell polarity and cellular cycling
between motile and dividing states for contact inhibition of
proliferation was explored using a disk model~\cite{SchMolYam2020}.
However, the question whether mechanical constraints alone, without
further intercellular interaction, can explain the observed
tissue-scale dynamics remains still largely open.

\begin{figure}
    \includegraphics[width=\linewidth]{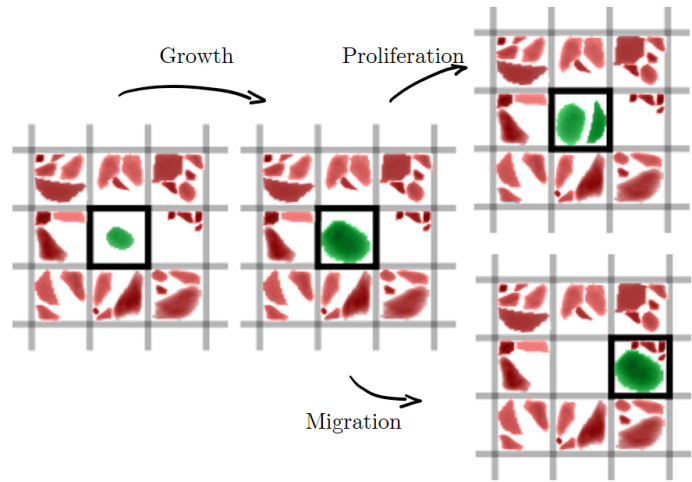}
    \caption{Illustration of the cellular automaton (CA) model. Cells
      (red and green) with individual sizes $A_i$ are positioned on a
      regular grid (gray $3\times3$ squares represent a section of
      this grid), where each square of the grid can be occupied by
      several cells as long as their total size does not exceed the
      size of the square $\sum_i A_i \leq A_{\text{max}}$. Each cell
      $i$ can proliferate with rate $\gamma(A_i)$, migrate with rate
      $\mu$, and grow logistically with rate $\alpha$ to a maximal
      size of $A_\text{max}$ (each action exemplarily sketched for the
      green highlighted cell in the central square). Cells interact
      via exclusion principle, i.e., migration and growth of
      individual cells is inhibited by lack of sufficient free space.
      Dynamics are numerically integrated using
      Markov-chain-Monte-Carlo method with time increment $\Delta t$.
      Note that the position of a cell is not resolved beyond its grid
      square.}
    \label{fig:developmentgraph}
  \end{figure}

We study this question by proposing a minimal model, an extension of
probabilistic cellular automata (CA) that allows describing
heterogeneous cell sizes and growth of individual cells. In our model,
cells located on the nodes of a 2D regular grid can grow, divide and
migrate, see \figref{developmentgraph} for an illustration. They only
interact by the exclusion principle, which means that a cellular
action is prevented if there is insufficient space. We calibrate the
model parameters, which quantify single-cell dynamics, to previous
in-vitro observations based on single-cell measurements from Puliafito
et al. (2012)~\cite{PulHufNevStrSigFygShr2012}, and subsequently
compare the emergent features of the calibrated CA model, such as
culture size and cell density, to independent experiments from the
same study. These features of CA model reproduce quantitatively the
corresponding experimental results, implying that the exclusion
principle, the only intercellular interaction incorporated in the
model, paired with size dependent proliferation rates is sufficient to
generate the observed contact inhibition. We thus demonstrate the
effectiveness of the CA model for efficient quantitative predictions
of tissue organization on the basis of rule-based cellular behavior
with heterogeneous cell sizes. We discuss the potential to incorporate
effective biomechanical interactions between cells in our CA model for
future studies of contact inhibition.

  \begin{figure*}
    \includegraphics[width=\linewidth]{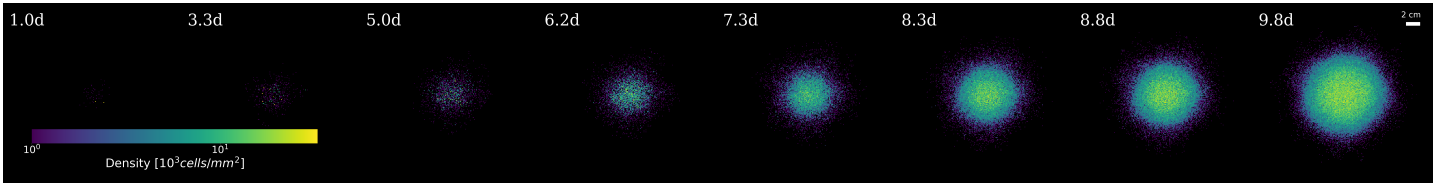}
    \caption{Exemplary simulation from the CA model of the growth of
      epithelial tissue in the confluent regime (1~d, 3.3~d, 5~d), as
      well as at (6.2~d, 7.3~d) and post-transition (8.3~d, 8.8~d,
      9.8~d) to a stationary postconfluent state. For each time point,
      the local cell density is encoded in color. The growing culture
      expands outwards over time. Around day 6 the density increases
      drastically in the center of the culture, as the lack of free
      space inhibits migration and cell growth, reflecting the
      transition to a postconfluent state. At the same time, a radial
      density gradient is visible, implying that cells at the boundary
      of the culture are able to grow larger and drive further
      outwards expansion. Model parameters calibrated as written in
      the main text. Note that the local average cell size is
      approximated by the displayed cell density as each pixel
      represents a lattice square of size
      $A_\text{max} = 990 \ \mu\text{m}^2$, e.g., a grid square with
      density of $10^1 \cdot 10^3 \ \text{cells/mm}^2$ hosts $9.9$ cells
      with average size $\leq$ 1/density $= 100 \ \mu\text{m}^2$.}
    \label{fig:timeseriesdevelopment}
\end{figure*}

\section{Data-driven approach}

Our data-driven approach is based on the mentioned experiments on the
development of epithelial tissue of MDCK
cells~\cite{PulHufNevStrSigFygShr2012}. In this study, the transition
from a confluent phase to a postconfluent phase is measured during the
expansion of the epithelial colony in terms of colony area, cell
density, average cell cycle time $\bar{\tau}_2$, and average velocity
of cell motion over time. Furthermore, it is observed that the cell
behavior in the center of the expanding colony at the end of the
confluent phase is similar to confluent, but homogeneously seeded cell
cultures, except that the transition to a stationary, postconfluent
state occurs soon after initiation of the culture. Thus, homogeneously
seeded cultures are exploited to further quantify the long-term
behavior at and after the transition from the confluent phase in terms
of the asymmetry $G(f)$ of daughter cell sizes after division,
distribution of cell sizes $\rho(A)$, cell size-dependent division
time $\tau_2(A)$, and cell median size over time.

In the following, we develop an according CA model
(\ref{sec:CA}.~\emph{CA model}), whose parameters are calibrated based
on the measured asymmetry $G(f)$, distribution $\rho(A)$, average cell
cycle time $\bar{\tau}_2$, cell size-dependent proliferation rate
$\gamma(A)=\ln(2)/\tau_2(A)$, and time point of the transition $t_T$
(\ref{sec:calibration}.~\emph{Parameter calibration}). The emergent
dynamics of the calibrated CA model are then successfully compared to
the experiment in terms of colony area, cell density, and average
velocity of cell motion over time for the setting of an expanding
colony as well as cell median size over time for the setting of
homogeneously seeded cells (\ref{sec:results}.~\emph{Results}). Thus,
the parameter calibration (based on single-cell dynamics) and
subsequent model assessment (based on emergent features of the tissue)
rely on separate datasets from independent experiments.

\section{CA model \label{sec:CA}}

We present an efficient, agent-based dynamical model to simulate the
development of two-dimensional tissue. The model is based on a
probabilistic cellular automaton (CA) but additionally allows for size
changes of individual cells during cell growth and division, see
\figref{developmentgraph} for an illustration of the model and
\figref{timeseriesdevelopment} for an exemplary simulation of
epithelial tissue. In the CA, the two-dimensional space is discretized
into a regular grid of $M^2$ squares each with size $A_{\text{max}}$,
which can be occupied by cells. Several cells $i$ with size $A_i$ can
occupy the same square as long as their cumulative size does not
exceed its size $\sum_i A_i \leq A_{\text{max}}$. While the position
of a cell is not resolved beyond its grid square, the restriction of
the total size ensures a monolayer without overlap of cells as in the
experiment. Cell dynamics are defined by actions with corresponding
rates, i.e., a cell proliferates with rate $\gamma$, migrates from one
grid square to an adjacent one with rate $\mu$, and grows with rate
$\tilde{\alpha}$. Further interactions could be incorporated, like
cell-cell adhesion~\cite{RosBoeLanVos2021,FraAlaBoeVosLan2022} or
cell-substrate adhesion, but are neglected to focus on a minimal
dynamical model.

The proliferation rate $\gamma(A)$ and growth rate $\tilde{\alpha}(A)$
are both assumed to depend on the size of the cell $A$ as suggested by
single-cell measurements~\cite{PulHufNevStrSigFygShr2012}, while the
migration rate $\mu$ is held constant. Cells interact with each other
solely via exclusion principle, that is a cell can only grow or
migrate if sufficient free space is available in its current grid
square or the adjacent ones. This means that the values
$\tilde{\alpha}(A)$ and $\mu$ represent the maximal possible rates for
growth and migration of a single cell of size $A$, whereas typically
these processes are mitigated or inhibited by neighboring cells
competing for the same free space.

The proposed model is a continuous-time cellular automaton, where
updates are asynchronous with time to next reaction being an
exponential random variable and its dynamics are as usually
numerically approximated by a Markov-chain-Monte-Carlo simulation with
Monte-Carlo-time-step $\Delta t$. Within a time step $\Delta t$, if
there are $N$ cells present, $3N$-times a random cell and a random
action - proliferation, migration or growth - with corresponding rate
$\omega$ are chosen. The cell then performs the chosen action with
probability $\omega \Delta t$. This procedure emulates numerically
that on average each cell performs each action with corresponding
rate, i.e., the probability that the action has not been performed
decreases exponentially $\sim \exp(-\omega t)$, given a sufficiently
small time step $\Delta t \ll \text{max}\{\gamma^{-1},\mu^{-1}\}$.

\subsection{Proliferation}

Proliferation is defined as a potentially asymmetric division of a mother
cell into two daughter cells keeping the overall area preserved.
Measurements of single-cell division times~\cite[Fig.
4E]{PulHufNevStrSigFygShr2012} suggest a sigmoidal function for the
dependence of the proliferation rate $\gamma(A)$ on the size $A$ of
the mother cell
\begin{equation}
    \gamma(A) = \gamma_{\text{max}} \exp \left[-
      \left(\frac{A_0}{A}\right)^m\right] \ .
    \label{eq:sizedependentproliferationmodel}
\end{equation}
This rate increases with cell size and becomes constant at large
$\gamma(A\gg A_0) \approx \gamma_{\text{max}}$ and small sizes
$\gamma(A\ll A_0) \approx 0$, while the steepness of the transition at
$A_0$ is set by the exponent $m$. Further, measurements of the cell
areas during single-cell division~\cite[Inset of Fig.
4C]{PulHufNevStrSigFygShr2012} suggest an asymmetry between the areas
of the two daughter cells, which is taken into account: If a mother
cell of size $A$ divides, one of the daughter cells is assigned a
fraction $f$ of the mother cell and while the other daughter is
assigned the remaining part, meaning the daughter cells have size $fA$
and $(1-f)A$. Naturally, the probability density $G(f)$ of the
fraction $f\in[0,1]$ is symmetric around $1/2$. Note that the total
cell area is not always conserved in the experiment~\cite[Fig.~4c,
lower inset]{PulHufNevStrSigFygShr2012}, which is ignored in the model
for simplicity.

\subsection{Migration}

For migration cells move with rate $\mu$ to a randomly selected
neighboring grid square. For grid square size $A_\text{max}$ this
implies a maximal migration velocity
$v_\text{max} = \mu \sqrt{A_\text{max}}$ across the grid or
\begin{equation}
    \mu = \frac{v_\text{max}}{\sqrt{A_\text{max}}} \ . \label{eq:migrate}
\end{equation}
Cell migration is only performed if the target grid square has
sufficient free space to accommodate the migrating cell.

\subsection{Growth}

We assume a generic, logistic cell growth
\begin{equation}
    \frac{\mathrm dA}{\mathrm dt} = \alpha A \left( 1 -
      \frac{A}{A_\text{max}} \right) \label{eq:cellgrowthmodel}
\end{equation}
with maximal cell size $A_\text{max}$ and growth rate $\alpha$. Note
that the growth rate $\alpha$ indirectly affects cell proliferation,
as it affects the position and width of the cell size distribution,
which in turn determines the average proliferation rate according to
\eqref{sizedependentproliferationmodel}. For instance, increasing the
growth rate $\alpha$ leads to accelerated growth of the cell culture.
While the logistic growth \eqref{cellgrowthmodel} is time-continuous
and could be integrated directly via forward-Euler-scheme or the
corresponding analytical solution, we employ for simplicity the same
Markov-chain-Monte-Carlo scheme as for proliferation and migration. We
use $\tilde{\alpha} = 1/\Delta t$ as the rate of the growth action,
meaning it is performed on average once per time step and cell, and
define the action as increasing the cell size by
$\Delta{}A = \alpha A ( 1 - A/A_\text{max}) \Delta t$. Note that the
time step is chosen sufficiently small such that
$\mathrm \Delta{}A/A \ll 1$. Furthermore, the increment $\Delta{}A$ is
limited to the available free space in the grid square. It should be
pointed out that in the previously proposed 1D vertex model, cell
growth was assumed to be a consequence of the surrounding tissue
stretching the cells~\cite{PulHufNevStrSigFygShr2012}. In contrast, we
explicitly choose a growth function that does neither rely on cells
actively pushing or pulling on each other, to devise a model with
minimal cell-cell interactions.

\section{Parameter calibration} \label{sec:calibration}

\begin{figure}
    \includegraphics{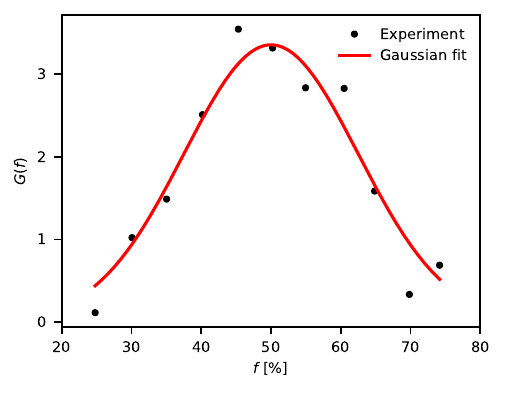}
    \caption{Asymmetry $G(f)$ of the daughter cell sizes after cell
      division is well described by a Gaussian distribution with
      expectation value $\bar{f} = 1/2$, standard deviation
      $\sigma = 1/8$, and a cut-off at large and small cell sizes
      $f\in[1/4,3/4]$, reflecting the absence of smaller or larger
      daughter cells (black dots represent data from Ref.~\cite[Fig.
      4C (inset)]{PulHufNevStrSigFygShr2012} obtained from tracking
      single-cell divisions, Gaussian distribution shown as red line).
      It is assumed that during division of a cell, one daughter cell
      has a fraction $f$ of the size of the mother cell, while the
      other cell has a fraction $(1-f)$.}
    \label{fig:division-asymmetry}
\end{figure}

The biologically motivated parameters $\gamma(A)$, $v_\text{max}$,
$\alpha$ of the CA model are calibrated based on in-vitro
observations, mostly of single-cell dynamics, during the growth of
normally differentiating epithelial cell cultures of Madin-Darby
canine kidney cells~\cite{PulHufNevStrSigFygShr2012}:
(\ref{sec:prolif}) The division asymmetry $G(f)$ is directly fitted to
data from tracking of single cell divisions. Then, this asymmetry
$G(f)$ is incorporated into a partial differential equation model to
optimize the proliferation rate $\gamma(A)$ based on cell size
distributions measured in cell tracking experiments. The obtained rate
$\gamma(A)$ is validated against independent data of cell division
times in a stationary, postconfluent state. (\ref{sec:migration}) The
maximal migration velocity $v_\text{max}$, and consequently the
migration rate $\mu$, is estimated from the time point of the
transition to a stationary, postconfluent state and the average cell
cycle time. (\ref{sec:growth}) Finally, the growth rate $\alpha$ is
optimized using simple Monte-Carlo-Simulations of independent cells
based on the average cell cycle time using the already calibrated
parameters $\gamma(A)$, $G(f)$. Note that the two mathematical models
supporting the calibration are much simpler than the CA model, without
spatial resolution or inter-cellular interaction.

\subsection{Proliferation \label{sec:prolif}}

Firstly, the probability density $G(f)$ of the division asymmetry can
be directly fitted to corresponding measurements from tracking
single-cell divisions, see \figref{division-asymmetry}. The data is
well described by a Gaussian distribution
\begin{equation}
  G(f) \sim \exp\left[-\frac{1}{2} \left(\frac{f-\bar{f}}{\sigma}\right)^2\right] \label{eq:Gp-symmetry-eqn}
\end{equation}
with expectation value $\bar{f}=\frac{1}{2}$, standard deviation
$\sigma=\frac{1}{8}$, boundaries of the fraction
$f \in [\frac{1}{4},\frac{3}{4}]$, and corresponding normalization
$\int_{\frac{1}{4}}^{\frac{3}{4}} G(f) \mathrm{d}f = 1$.

\begin{figure}[!hb]
  \includegraphics{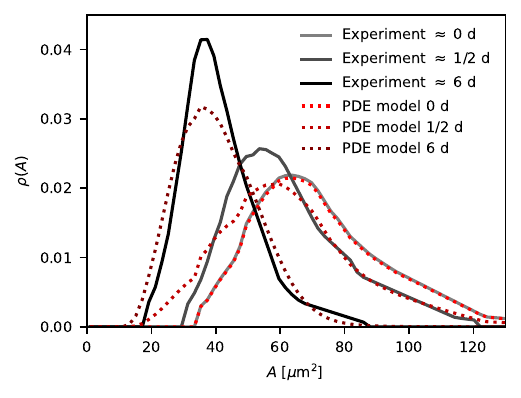}
  \caption{Distribution of cell sizes over time after transition to a
    stationary, postconfluent state (solid lines selected from
    experimental data~\cite[Fig. 4D]{PulHufNevStrSigFygShr2012}) is
    well described by partial differential equation (PDE
    \eqref{prolif-model}, dashed lines), which only incorporates cell
    size-dependent proliferation $\gamma(A)$,
    \eqref{sizedependentproliferationmodel}, and asymmetric cell
    division $G(f)$, \eqref{Gp-symmetry-eqn}. The distribution at day
    0 serves as initial condition for the PDE. Deviations from the
    experimental distributions originate mainly from overestimated
    tails at small cell sizes, which could hint to modification of the
    asymmetry of the cell division at cell sizes
    $<100 \ \mu{}\text{m}^2$, see main text.}
    \label{fig:A-distribution}
\end{figure}

The parameters $\gamma_\text{max}$, $A_0$, $m$ defining the
proliferation rate $\gamma(A)$ are optimized based on the observed
dynamics of the cell size distribution $\rho(A,t)$, see solid lines in
\figref{A-distribution}. These distributions are measured in a regime
of the cell culture growth, where cells no longer migrate or increase
their size due to their dense packing. Instead, these cells only
divide into smaller and smaller cells whose proliferation rates
decrease with their size. We model this loss of larger cells and rise
of smaller cells in the distribution $\rho(A)$ using a partial
integro-differential equation (PDE)
\begin{equation}
    \label{eq:prolif-model}
    \begin{aligned}
        \frac{\partial \rho}{\partial t}(A,t) = &- \gamma(A)\rho(A,t)
        + \int \mathrm df \ G(f) \ \gamma \left(\frac{A}{f}\right) \rho\left(\frac{A}{f},t\right)\\
        &+ \int \mathrm df \ G(f) \ \gamma \left(\frac{A}{1-f}\right)
        \rho\left(\frac{A}{1-f},t\right) \\
                \underset{G(f)=G(1-f)}{=} &- \gamma(A)\rho(A,t) +  2 \int \mathrm df \ G(f) \ \gamma \left(\frac{A}{f}\right)
        \rho\left(\frac{A}{f},t\right)\ .
    \end{aligned}
\end{equation}
The first term reflects the loss of cells with size $A$ due to
division with proliferation rate $\gamma(A)$. The second and third
term reflect the gain of cells with size $A$ due to division of larger
cells with sizes $A/f$ or $A/(1-f)$, where one of the resulting
daughter cells obtains a fraction $f$ or $1-f$, respectively, of the
mother's size. Note that \eqref{prolif-model} simplifies to the
equation proposed in Ref.~\cite{PulHufNevStrSigFygShr2012}, if a delta
distribution $G(f) = \delta(f-1/2)$ is assumed. We solve
\eqref{prolif-model} numerically by discretizing the cell size
$A \to A_{i\in \mathbb{N}}$ and integrating the resulting set of
ordinary equations with {\tt odeint}~\cite{Viretal2020}. We optimize
the parameters $\gamma_\text{max}$, $A_0$, $m$ of the proliferation
rate $\gamma(A)$ by minimizing the distance $\mathcal{D}$ between the
distributions $\rho_\text{exp}(A_i)$, $\rho_\text{PDE}(A_i)$ obtained
from experiment and PDE, respectively, at selected times $t$
\begin{equation}
  \mathcal{D} = \sum_{t} \sum_{i\in \mathbb{N}}
    \left[\rho_\text{exp}(A_i, t) - \rho_\text{PDE}(A_i, t, \gamma_\text{max}, A_0, m)\right]^2 \label{eq:l2err}
\end{equation}
using the python package {\tt scipy.optimize}~\cite{Viretal2020}. We
select experimental distributions at three characteristic time points
from the experiment, where the distribution at the first time point
$\rho_\text{exp}(A_i,t\approx0)$ serves as initial condition for the
PDE while the other two time points $t\approx 0.5$~d and
$t\approx 6$~d are the reference points for the optimization in
\eqref{l2err}. While distributions are reported at more intermediate
time points in Ref.~\cite{PulHufNevStrSigFygShr2012}, they do not
significantly contribute to the optimization as these distributions
transition smoothly into each other and thus can be neglected for
clarity.

\begin{figure}
  \includegraphics{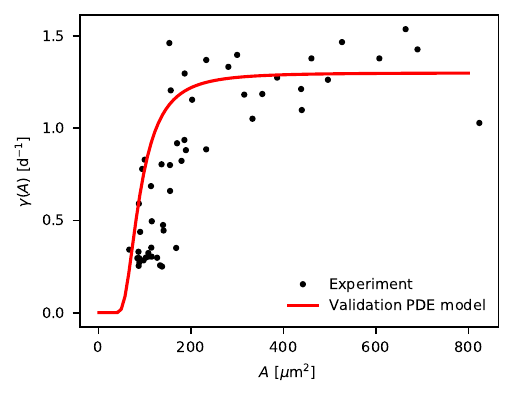}
  \caption{Cell size-dependent proliferation rate $\gamma(A)$ (red
    line, \eqref{sizedependentproliferationmodel} with
    $\gamma_\text{max} = 1.3 \ \text{d}^{-1}$,
    $A_0 = 80 \text{ }\mu{}\text{m}^2$, $m = 3$) optimized with PDE
    model based on cell area distributions after transition to a
    stationary, postconfluent state, see \figref{A-distribution}, is
    consistent with cell division times $\tau_2\approx \ln(2)/\gamma$
    (black dots from single-cell tracking~\cite[Fig.
    4E]{PulHufNevStrSigFygShr2012}). In particular, proliferation
    rates decrease rapidly for cell areas below
    $A<200 \mu{}\text{m}^2$. }
  \label{fig:proliferation-rate}
\end{figure}

We obtain parameters $\gamma_\text{max} \approx 1.3 \ \text{d}^{-1}$,
$A_0 \approx 80 \text{ }\mu{}\text{m}^2$, $m \approx 3$, for which we
observe good agreement both with the dynamics of the cell size
distributions, see \figref{A-distribution}, and the division times of
single cells, see \figref{proliferation-rate}. Deviations from the
experimental distributions in \figref{A-distribution} originate mainly
from overestimated tails at small cell sizes, left of the maximum of
each distribution. These deviations seem to be diminished, when the
domain of the size fraction $p$, \eqref{Gp-symmetry-eqn}, is further
narrowed around the symmetric case $p=1/2$. Thus, the deviations in
\figref{A-distribution} could hint at a narrowing of the asymmetry of
the cell division at small at cell sizes $<100 \ \mu{}\text{m}^2$,
e.g., due to a minimal absolute size a cell must have. In fact, the
distribution $G(p)$, which the PDE relies on, has been obtained for
mother cell sizes $A\in[100,1000] \ \mu\text{m}^2$, due to the
difficulty of tracking cells below $70 \ \mu\text{m}^2$, while the
cell size distributions $\rho(A)$ were mostly measured at
$A<100 \mu{}m^2$~\cite{PulHufNevStrSigFygShr2012}. Furthermore, the
PDE assumes conservation of total cell area for simplicity, which is
not always the case in the experiment~\cite[Fig.~4c, lower
inset]{PulHufNevStrSigFygShr2012}.

\subsection{Migration \label{sec:migration}}

According to \eqref{migrate}, the migration rate $\mu$ is directly
defined by the maximal migration velocity $v_\text{max}$. This
velocity can be estimated from the transition point $t_T$ from
exponential to subexponential growth of the cell culture (transition
to a stationary, postconfluent
state~\cite{PulHufNevStrSigFygShr2012}): As cells proliferate and grow
in size, they inevitably migrate outwards causing the cell culture to
expand. As long as this expansion is fast enough, each cell has
sufficient space to grow and divide at maximal capacity and the cell
culture grows exponentially, while the density in the culture remains
constant, see black points in Figs.~\ref{fig:colony-area}
and~\ref{fig:colony-density} before $t<5$~d, respectively. However, at
some point the outward migration of the cells cannot compensate for
the increase in total cell size and consequentially the growth of
invidiual cells in the center of the culture is inhibited by the lack
of free space. This leads to subexponential growth of the culture and
an increase in cell density, see black points in
Figs.~\ref{fig:colony-area} and~\ref{fig:colony-density} after
$t>6$~d, respectively. For an exponential growth of a circular cell
culture
\begin{equation}
  A_\text{cult}(t) = A_\text{cult}(0) \exp(\bar{\gamma} t) = \pi R^2_\text{cult}(t)
\end{equation}
with the average proliferation rate
$\bar{\gamma} = \ln(2)/\bar{\tau}_2$, given in terms of the reported
average cell cycle time
$\bar{\tau}_2 = 0.75 \pm 0.14$~d~\cite{PulHufNevStrSigFygShr2012}.
Note that in the context of the model, the rate $\bar{\gamma}$ should
reflect the average proliferation rate over an ensemble of
independently proliferating, \eqref{sizedependentproliferationmodel},
and growing, \eqref{cellgrowthmodel}, cells, which is latter exploited
for the calibration of (\ref{sec:growth}) cell growth. Along with the
initial culture size
$A_\text{cult}(0) \approx 1.8\cdot10^4 \mu\text{m}^2$, the maximal
migration velocity $v_\text{max}$ can then be estimated from the
condition
\begin{equation}
\begin{split}
  v_\text{max} &= \left. \frac{\mathrm d R_\text{cult}}{\mathrm d t}
  \right|_{t_T} = \left.\frac{\mathrm d R_\text{cult}}{\mathrm d
    A_\text{cult}} \cdot \frac{\mathrm d A_\text{cult}}{\mathrm d
    t}
  \right|_{t_T} \\ &
  = \frac{\bar{\gamma}}{2} \sqrt{\frac{A_\text{cult}(0)}{\pi}}
  \exp\left(\frac{\bar{\gamma}}{2} t_T\right) \ . \\
\end{split}
\label{eq:condition}
\end{equation}
The range of the cell cycle time $\bar{\tau}_2$ and the transition
point $t_T = 6 \pm 1$~d imply a range for the maximal migration
velocity $v_\text{max} \in [9,100] \ \mu\text{m}/\text{h}$. In
Ref.~\cite{PulHufNevStrSigFygShr2012} a coarse consistency check
between the observed average cell velocity, transition time $t_T$,
average cell cycle time $\bar{\tau}_2$, and colony area
$A_\text{cult}(t_T)$ was performed based on the same assumption as
\eqref{condition}. However, the ranges for $t_T$ and $\bar{\tau}_2$
were neglected and the maximal and average cell velocity were assumed
to be the same, while the former is a single-cell parameter and the
latter an emergent feature from the interaction with other cells. Note
that the condition in \eqref{condition} is a simplification and rather
estimates a lower limit for $v_\text{max}$. The transition from
exponential to subexponential growth is expected before the rim of the
cell culture actually reaches an outward velocity $v_\text{max}$. For
instance, at $t=5$~d the culture already exhibits deviations from
exponential growth, while according to \eqref{condition} the velocity
of the boundary at this point is only $8-30 \ \mu$m. In order to
reflect the estimated range for $v_\text{max}$, we use its upper limit
$v_\text{max} = 100 \ \mu \text{m}/\text{h}$ in the following, but
also report the results for
$v_\text{max} = 50 \ \mu \text{m}/\text{h}$ in SI
Figs.~\ref{fig:app-colony-area}-\ref{fig:app-median}.

\subsection{Cell growth \label{sec:growth}}

The average proliferation rate $\bar{\gamma}$ derived from the
measured average cell cycle time is also used to estimate the cell
growth rate $\alpha$: Firstly, the maximal cell size is set to
$A_\text{max} = 990 \ \mu\text{m}^2$ based on tracking of sizes of
single cells~\cite[Fig. 4A]{PulHufNevStrSigFygShr2012} and the
observed minimal cell density. For the phase of exponential growth of
the culture, it is reasonable to assume that cells proliferate and
grow independently of each other according to
Eqs.~(\ref{eq:sizedependentproliferationmodel}),
(\ref{eq:cellgrowthmodel}), (\ref{eq:Gp-symmetry-eqn}). We compute
these simple dynamics numerically for ensembles of cells using
Monte-Carlo simulations. In particular, we use the parameters of
proliferation $\gamma(A)$, $G(p)$ as calibrated above and additionally
optimize the growth rate $\alpha$ of logistic cell growth
\eqref{cellgrowthmodel} such that the resulting exponential slope of
the total size of the cell ensemble matches with the slope
$\bar{\gamma} = \ln(2)/\bar{\tau}_2$ corresponding to the average cell
cycle time $\bar{\tau}_2 = 0.75 \pm 0.14$~d measured independently by
single-cell tracking. From this optimization a growth rate
$\alpha = 1.3 \pm 0.3 \ \text{d}^{-1}$ is obtained.

\subsection{Initial and boundary conditions}

For expanding culture experiments, the size of the spatial grid is
chosen sufficiently large, such that cells do not reach the boundary
of the grid during simulation. We obtain the initial condition of the
CA model by starting from a single cell of maximal size in the center
of the grid and then simulating the dynamics until the size of the
culture matches the initial one in the experiment. For experiments
where cells are seeded homogeneously, a $12 \times 12$-grid with
reflecting boundary conditions is used and as initial condition a
single cell of maximal size is placed in every second grid point,
reflecting the experimental seeding density of $600$~cells/mm$^2$.
Finally, the time increment in the CA is set sufficiently small
$\Delta t = 6$~min such that
$\Delta t \ll \text{max}\{\gamma_\text{max}^{-1},\mu^{-1}\}$ and
$\mathrm \Delta{}A/A \ll 1$.

\section{Results \label{sec:results}}

\begin{figure}
  \includegraphics{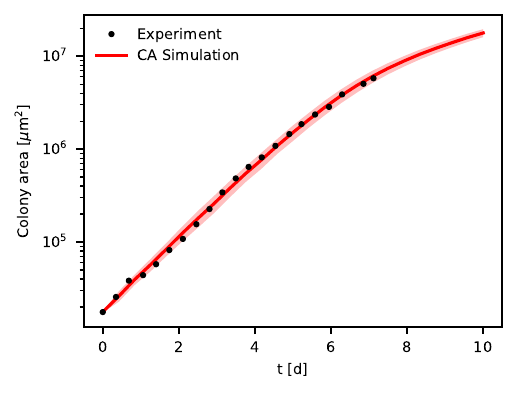}
  \caption{Emergent dynamics of calibrated CA model (red line) reproduces
    experimentally observed size of expanding epithelial colony (black
    dots from Ref.~\cite[Fig. 1B]{PulHufNevStrSigFygShr2012}). The
    colony size exhibits initially an exponential growth (low-density,
    confluent regime) and slows around time $t_T = 6 \pm 1$~d to a
    subexponential growth (transition to a stationary, postconfluent state).
    Results from CA model are averaged over $15$ independent runs and
    standard deviation is displayed as red shaded region. Model
    parameters calibrated as written in the main text.}
    \label{fig:colony-area}
\end{figure}

\begin{figure}[t]
  \includegraphics{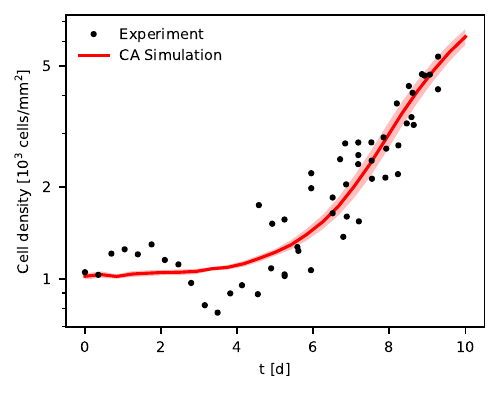}
  \caption{Emergent dynamics of calibrated CA model (red line)
    reproduces experimentally observed cell densities of expanding
    epithelial colony (black dots from Ref.~\cite[Fig.
    1C]{PulHufNevStrSigFygShr2012}, corresponding data to areas
    displayed in \figref{colony-area}). The cell density is initially
    constant (low-density, confluent regime) and increases rapidly
    around time $t_T = 6 \pm 1$~d (transition to a stationary,
    postconfluent state) implying dense packing of cells, which become
    smaller and smaller due to slowing but continued proliferation
    without cell growth. Note that the small variations ($\pm 30\%$)
    until day 6 appear larger due to the logarithmic y-scale and were
    interpreted as constant density by the original
    experimenter~\cite{PulHufNevStrSigFygShr2012}. Results from CA
    model are averaged over $15$ independent runs and standard
    deviation is displayed as red shaded region. Model parameters
    calibrated as written in the main text.}
    \label{fig:colony-density}
\end{figure}

\begin{figure}
  \includegraphics{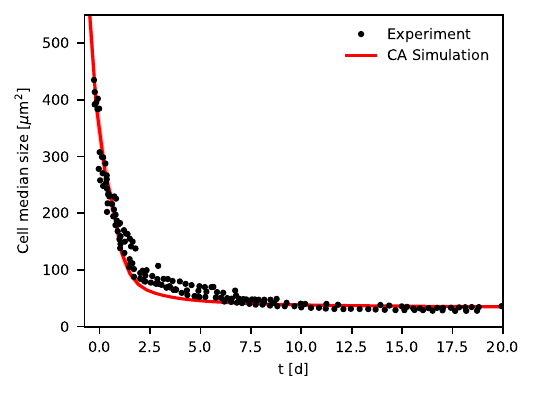}
  \caption{Emergent dynamics of calibrated CA model (red line)
    consistent with experimentally observed cell size median for
    initially homogeneously seeded cells (black dots from
    Ref.~\cite[Fig. 3C]{PulHufNevStrSigFygShr2012}, transition $t_T$
    to a stationary, postconfluent state corresponds roughly to
    $1.5$~d). Cell size median initially drops rapidly and then
    saturates at small cell sizes $A\sim 40 \ \mu\text{m}^2$. Results
    from CA model are averaged over $15$ independent runs and standard
    deviation is displayed as red shaded region.}
    \label{fig:median}
\end{figure}

\begin{figure}[t]
  \includegraphics{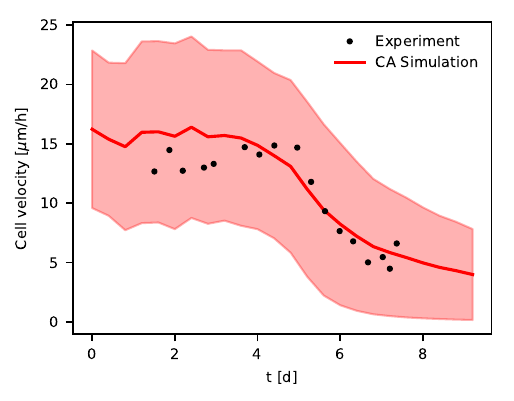}
  \caption{Emergent dynamics of calibrated CA model (red line)
    reproduces qualitatively the experimentally observed cell
    velocities of expanding epithelial colony (black dots from
    Ref.~\cite[Fig. 2C]{PulHufNevStrSigFygShr2012}, corresponding data
    to Figs.~\ref{fig:colony-area} and~\ref{fig:colony-density}). The
    average cell velocity is initially constant (low-density,
    confluent regime) and decreases rapidly around time
    $t_T = 6 \pm 1$~d (transition to a stationary, postconfluent
    state). Presented are the root-mean-square velocities, where the
    model velocities are computed from a single simulation run over
    time increments of $580$~min to neglect small-scale oscillations.
    Due to the presence of a radial velocity gradient, see
    \figref{direction}, the corresponding standard deviation of the
    velocities (red shaded region) is quite large.}
    \label{fig:cell-velocity}
\end{figure}

\begin{figure*}
  \centering
    \includegraphics{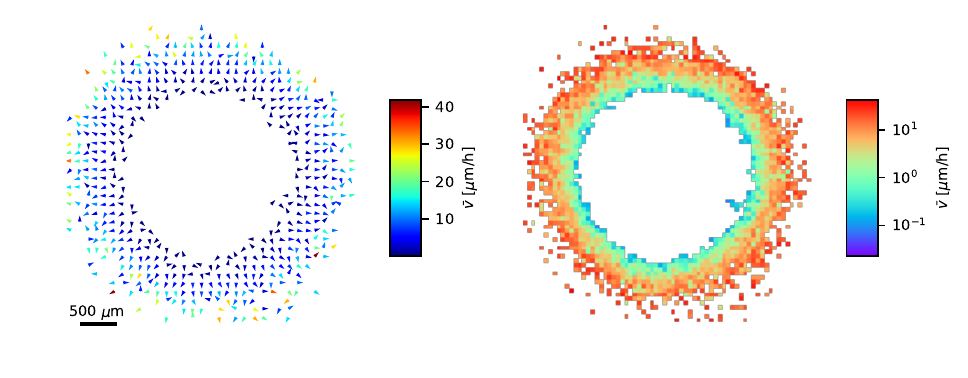}
    \caption{CA model displays an outwards biased cell motion (right
      panel) accompanied by a velocity gradient from inner region of
      colony to its periphery (left and right panel encoded by color).
      Displayed are cell velocities $\bar{v}$ of the last $580$~min of
      the simulation (posttransition) spatially averaged (left panel:
      $4\times4$ grid squares, right panel: higher resolution
      $2\times2$ grid squares) where arrows represent direction of
      motion (left panel) and color magnitude of velocity (linear
      scale in left panel and logarithmic scale in right panel for
      visibility, missing arrows or white squares mean no measurable
      cell motion). For a radial profile of the velocities see SI
      Fig.~\ref{fig:radial-profile}.}
    \label{fig:direction}
\end{figure*}

We find quantitative agreement between the calibrated CA model and
independent experiments with respect to several emerging features of
the epithelial cell culture. This validates our minimal model and
showcases its power to predict emergent tissue wide dynamics from
single cell behavior. The agreement also suggests that the exclusion
principle as sole inter-cellular interaction paired with a
size-dependent proliferation rate of each cell is sufficient to
explain the emerging dynamics of contact inhibition:

We implement the previously calibrated parameters into the CA model
and compare the emerging dynamics of culture size and cell density
with the experiment of epithelial colony growth, see
Figs.~\ref{fig:colony-area} and~\ref{fig:colony-density}. In the CA,
the culture size is computed as the total area of all grid points
containing at least a single cell and the cell density is computed as
ratio between this area and the total number of cells. The resulting
dynamics of culture size and cell density match with the experimental
ones. In particular, the CA model does not only quantitatively
reproduce the exponential growth and constant density in the initial
low-density confluent regime and the time point of the transition to a
stationary, postconfluent state, but also the curve at and after this
transition. The predictions of the CA model are also consistent for
the case of initially homogeneously seeded cells, see \figref{median},
which reflects the regime directly at and long after the transition to
a stationary, postconfluent state. Furthermore, the transition is
accompanied by a rapid decrease in the emergent cell velocities, which
are comparable between CA model and experiment, see
\figref{cell-velocity}. In particular, there is a velocity gradient
from the inside to the outside of the colony and an outwardly biased
flow, see \figref{direction} as well as SI
Figs.~\ref{fig:app-direction} and \ref{fig:radial-profile}.

Note that the calibration of the model parameters is only based on
observations from experiments independent from the one we compare the
model predictions to. The only exception is the time point of
transition $t_T$ used for the calibration of the maximal velocity
$v_\text{max}$, as the maximal velocity at the boundary of the culture
has not been directly measured. Apart from that, the majority of the
calibration is based on tracking of the dynamics of single cells and
the model parameters themselves describe only single-cell behavior.

The match of the emergent features, i.e., dynamics of culture size and
cell density, between experiment and CA model suggests that they are
determined by the single-cell dynamics. In particular, the CA model
does not incorporate any inter-cellular interactions, besides the
exclusion principle. In the CA model, the transition of the dynamics
at $t_T$ originates solely from the limitation that the boundary of
the culture cannot expand faster outwards than the maximal migration
velocity $v_\text{max}$ of the cells. Beyond the corresponding
critical size of the culture, cells in the center of the culture no
longer get access to the necessary space to grow, i.e.
$\mathrm dA/\mathrm dt \to 0$, and consequently become smaller due to
continued proliferation. The latter gets slower due to the
size-dependent, sigmoidal proliferation rate $\gamma(A)$. The capacity
of cells to grow separates their proliferation behavior into a pre-
($A\gg A_0 \to \gamma(A)=\gamma_\text{max}$) and posttransition regime
($A \sim A_0 \to \gamma(A)\ll\gamma_\text{max}$) according to
\eqref{sizedependentproliferationmodel}. Note that based on the
assumption that for times $t \gg t_T$ effectively only cells in a
boundary layer with thickness $\Delta R$ of the culture can
proliferate with rate $\bar{\gamma}$, the asymptotic, posttransition
growth may be estimated by
$\mathrm dA/\mathrm dt \approx 2\sqrt{\pi}\bar{\gamma}\Delta R \cdot
\sqrt{A} = \kappa \sqrt{A}$ or
$A(t\gg t_T) = \kappa^2/4 \cdot (t - t_0)^2$, analogous to recent
estimates for spherical cultures~\cite{FraMicAlaKunVosLan2023}, while
$\Delta R(v_\text{max})$ may depend on the maximal migration velocity.
We obtain an effective thickness of the proliferating rim
$\Delta R = 370 \pm 10 \ \mu$m ($t_0 = 3.3 \pm 0.2$~d) from the CA
simulation, consistent with rescaled results from a previous
theoretical model of epithelial colonies~\cite{AlaHatLowVoi2015}. Note
that while the cell size-dependent division times $t_2$ exhibited by
the calibrated CA model closely resemble the experimental ones, see SI
Fig.~\ref{fig:division-times}, the distribution of these times is
broader than in the experiment, see inset in
Fig.~\ref{fig:division-times}. In fact, the division times in the CA
model follow an exponential distribution by design for mathematical
and numerical convenience, whereas biological cells divide at more
regular intervals leading typically to a narrower Erlang-like
distribution. However, our results imply that this difference in the
cell cycle distributions does not affect the emergent dynamics of the
colony and the implementation of cell division according to regular
cell cycles into the CA model is straightforward.

In the CA model, the observed rapid decrease in cell velocity at the
transition results from cells in the inner region of the colony
slowing down due to insufficient space to migrate, which generates a
velocity gradient from inside to outside, see \figref{direction}. At
the same time cell motion transitions from undirected to a collective
outwards bias, see also \figref{app-direction}. This outwards directed
motion is an emergent feature of the interplay between the migration
rate, the exclusion principle, and the fact that more free space is
available outwards. It should be emphasized that for the CA model the
maximal migration velocity $v_\text{max}$ is significantly bigger than
the velocities that are measured for individual cells inside the
culture, see \figref{cell-velocity}, due to the exclusion principle.
For instance, 1 to 3 occupied neighboring grid squares reduce the
actual migration velocity by a factor of $3/4$ to $1/4$. Moreover, the
migration velocity in the CA model represents the speed of
instantaneous, undirected movement of single cells, whereas in
in-vitro experiments typically collective, directed movement of groups
of cells is measured, e.g., directed outwards in case of an expanding
culture. Consequently, this directed motion exhibits much smaller
velocities, consistent with previous experimental measurements of cell
velocity in epithelial tissue $\sim 25 \mu \text{m}/\text{h}$
depending on the setting (wound-cut technique, expanding colony,
...)~\cite{PulHufNevStrSigFygShr2012,TauChuSaiSat1979,RosMis1980,PetRefGraPouLadBugSil2010,JenPedMorParPedNej2019}.
Note that the exact value of the average velocity depends to some
degree on the chosen definition, i.e., the chosen time increment
between two cell position measurements and whether the average is
taken over the whole colony or a specific location. In
Ref.~\cite{PulHufNevStrSigFygShr2012} the cell velocity was extracted
from four local $450\times336 \ \mu\text{m}^2$ fields of view, while
for the CA model it is possible to average over the entire cell
culture. This explains the large range of velocities observed in the
CA model, which reflects the radial velocity gradient, see
\figref{direction}, while the standard error for the experimental
velocities is not reported. Nevertheless, we can infer from the match
of simulation and experiment in \figref{cell-velocity} that the course
of the cell velocities, initially constant with a rapid decrease
around the transition point, is reproduced by the CA model.

\section{Discussion}

We developed a mechanistic, data-driven model to simulate the dynamics
of planar cell cultures by extending a probabilistic cellular
automaton to incorporate size changes of individual cells during
growth and cell division. We successfully apply this CA model to
previous in-vitro experiments~\cite{PulHufNevStrSigFygShr2012} of
epithelial tissue composed of Madin-Darby canine kidney cells. In
particular, the model parameters are calibrated using measurements
based on single-cell tracking and subsequently the resulting CA model
is validated against independent experiments on the development of
culture size, cell density, and cell velocity. The agreement between
model predictions and these experimental observations suggests that
the exclusion principle as sole inter-cellular interaction paired with
a size-dependent proliferation rate of each cell is sufficient to
explain the emerging dynamics of contact inhibition. In particular,
incorporation of additional mechanisms like cell-cell signaling or
cell adhesion are not necessary. In the CA model, the reproduced
transition from exponential to quadratic growth of the colony
originates from the increasing lack of space in the center of the
colony, inhibiting cell growth and division, when the expansion of the
colony can no longer be compensated for by the migration of the
exterior cells. This origin of the transition is consistent with
previous interpretations and the observation that contact inhibition
can take place several days after cells have been in
contact~\cite{PulHufNevStrSigFygShr2012,AlaHatLowVoi2015}. Moreover,
the collective, outwards motion of the exterior cells and the rapidly
increasing cell density in the center of the culture are a consequence
of these simple mechanical constraints.

As mentioned there is a wide variety of theoretical models to describe
and simulate the development of healthy and cancerous cellular tissue
and even contact inhibition~\cite{GaoMcDHlaEnd2013, AndCha2012,
  SciPreWol2013, SciPre2012, SotVerTav2011, GivSciPreBuoFun2010,
  RejAnd2011, AhmGra2018, NobBurLeLemVioKatBee2022, PodDelElg2015}.
Most of these models explicitly consider intra- and inter-cellular
forces, allowing to study the delicate interplay between biomechanical
quantities like tension, pressure, adhesion, as well as cellular
stiffness and compressibility, which govern cell topology, e.g., the
distribution of the number of direct cell neighbors, and large scale
phenomena like elastic, plastic, viscoelastic deformations. For
example, previously a mechanistic hybrid continuum-discrete model has
been applied to the same experiment of epithelial cell colonies as our
study~\cite{AlaHatLowVoi2015}. Besides culture size, cell density and
cell area median, this model was able to reproduce the observed cell
topology and radial distribution function. However, the spatial scale
of the culture had to be fitted empirically to the culture size, as
the model underpredicts the cluster size where the transition to a
stationary, postconfluent state occurs, and the resulting cell
velocities are reported to be ten times smaller than in the
experiment. This is partially due to computational cost, which
prevented the use of a larger parameter of mobility, which also
reflects effects of cell-substrate adhesion. This example highlights
two more general issues of the above-mentioned types of models, see
also Ref.~\cite{Vos2012}: Firstly, they often rely on empirical or
effective parameters, like coefficients of free energy functions or
elastic properties of tissue, which are either biologically not known
or at least very difficult to measure, in contrast to the properties
of single cells. Secondly, they often entail high computational
effort, which limits exploration of the parameter space and
consequently the calibration to concrete experimental data.

While CA models have the advantage, that their parameters are usually
based on biologically motivated, single-cell behavior and that they
are computationally relatively cheap, the incorporation of
biomechanical forces has been a challenge with few exceptions like
cell adhesion~\cite{FraAlaBoeVosLan2022,RosBoeLanVos2021}. The type of
CA we propose here has the potential to partially overcome this
challenge: The fact that cells of different sizes can be located at
the same grid point naturally defines local cell densities and density
gradients. Together with the possibility to define cellular,
size-dependent mechanisms and forces, beyond proliferation and growth,
this allows to effectively introduce biomechanical processes from
other types of models, e.g., homeostatic
pressure~\cite{PodDelElg2015,BasRisJoaSasPro2009} or cell bulk stress
in terms of cell compression~\cite{AlaHatLowVoi2015} from the ratio of
the actual size of a cell and its (dynamic) target size. Note that the
concept of local densities has already been exploited in previous
spatial stochastic models, which allow multiple particles in each
lattice site~\cite{BarMan1996,SmiGri2016}, e.g., to define migration
probabilities based on the density differences between lattice sites,
which is similar to our CA model if cell size and cell growth is
omitted. Mean-field approximations of these models lead to partial
differential equations of the reaction-advection-diffusion type
exhibiting matching dynamics~\cite{SmiGri2016,CiaSmiGri2017}. However,
these models neglect cell growth, a crucial effect which complicates
the derivation of a corresponding continuum model. An example is the
already mentioned hybrid continuum-discrete model, derived from a
stochastic particle model using the framework of dynamic density
functional theory~\cite{AlaHatLowVoi2015}.

Furthermore, in our CA model cells are capable of moving past each
other, which is typically not inherent to both common CA models and
vertex models. This feature can be exploited to study effects like
collective motion or fingering~\cite{SzaMer2013,SzaVarGarHegCzi2012}
by adding direction vectors of persistent motion to each cell, which
are affected by neighboring cell motion or cell densities. This could
help to reproduce the finger-like protrusions observed for the
expanding epithelial colony~\cite{PulHufNevStrSigFygShr2012}, which
neither our CA model nor the corresponding vertex
model~\cite{AlaHatLowVoi2015} exhibit. Moreover, the concept of local
cell densities allows to couple the CA model to additional PDEs to
incorporate the interaction of cells with external fields like oxygen,
nutrients, and metabolic waste. Finally, the CA model can be directly
extended to three-dimensional space. However, features of small
spatial scales like cell shape~\cite{AlaHatLowVoi2015} and correlation
lengths~\cite{AlaHatLowVoi2015,PulHufNevStrSigFygShr2012} can not be
captured by our model. While none of these concepts were in the end
necessary for the application of the CA model to the considered
experiments of epithelial growth, they may guide efficient,
data-driven modeling of future scenarios of tissue dynamics.

\section{Acknowledgement}

We are grateful for discussions with S.~R\"uhle, M.~Luft and
F.~Freier. The authors acknowledge that this research has been
co-financed by the EU, the European Social Fund (ESF), and by tax
funds on the basis of the budget passed by the Saxon state parliament
(project SAB-Nr. 100382145), the Bundesministerium f\"ur Bildung und
Forschung (BMBF~16dkwn001a/b), and the S\"achsisches Staatsministerium
f\"ur Wissenschaft und Kunst (SMWK) (project FORZUG~II~TP~3). The
publication of this article is funded by the Open Access Publication
Fund of Hochschule f\"ur Technik und Wirtschaft Dresden - University
of Applied Sciences. The funders had no role in study design, data
collection and analysis, decision to publish, or preparation of the
manuscript.

\section{Conflicts of Interest}

The authors declare no conflict of interest.

\section{Data Availability}

All relevant data are within the manuscript and its Supporting
information files. The code is publicly available under
\url{https://github.com/j-schmied/cellularautomata}.

\setcounter{figure}{0}
\appendix
\section{Supplemental figures}

\begin{figure*}
    \includegraphics{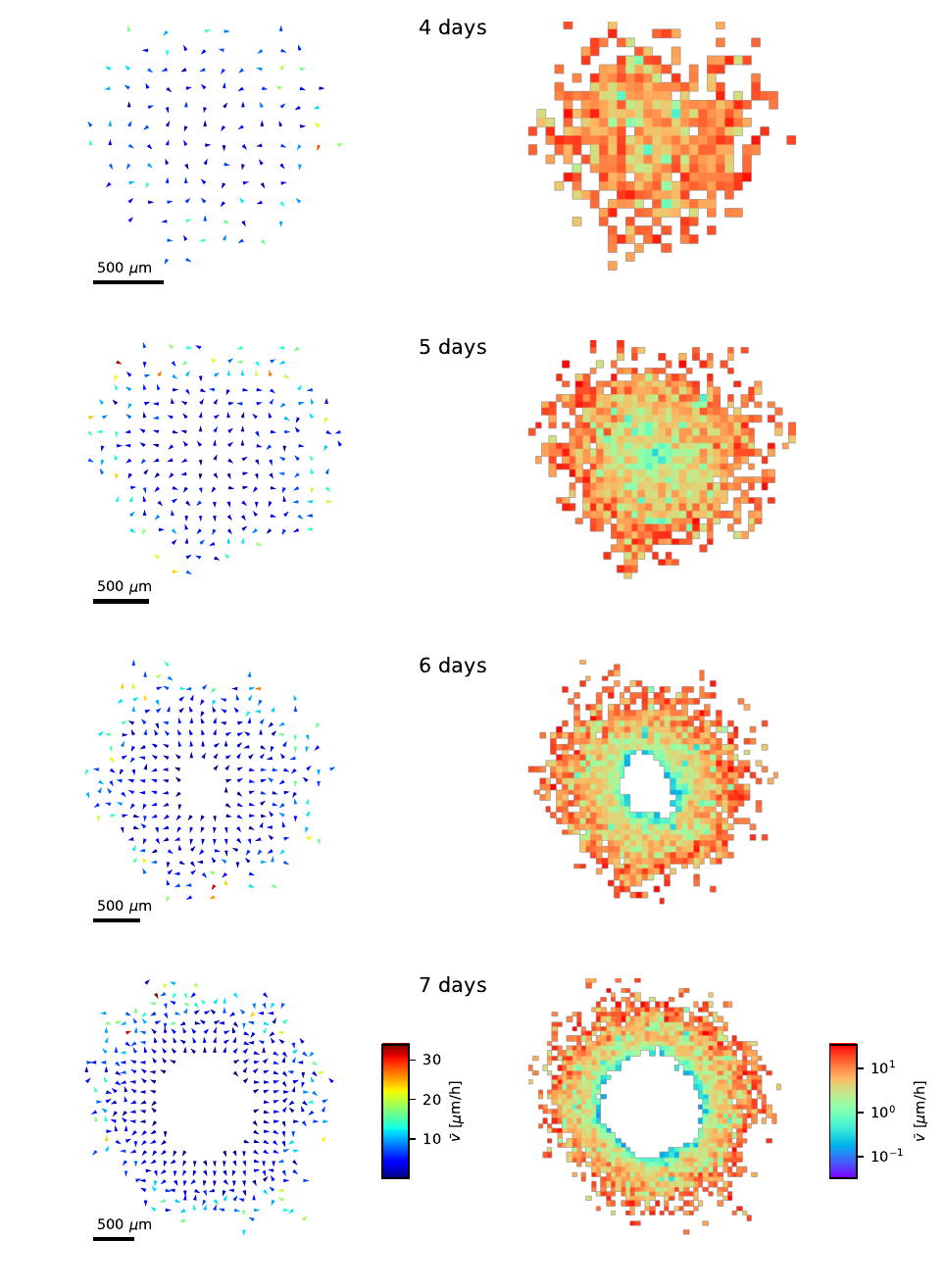}
    \caption{CA model displays transition from undirected, uniform,
      fast cell motion to outwards biased motion with radial velocity
      gradient (from top to bottom, note shrinking scale bar). Panels
      analogous to \figref{direction} in the main text, but for
      different times around the transition point $t_T=6 \pm 1$~d. For
      a radial profile of the velocities see
      Fig.~\ref{fig:radial-profile}.}
    \label{fig:app-direction}
\end{figure*}

\begin{figure}
    \includegraphics{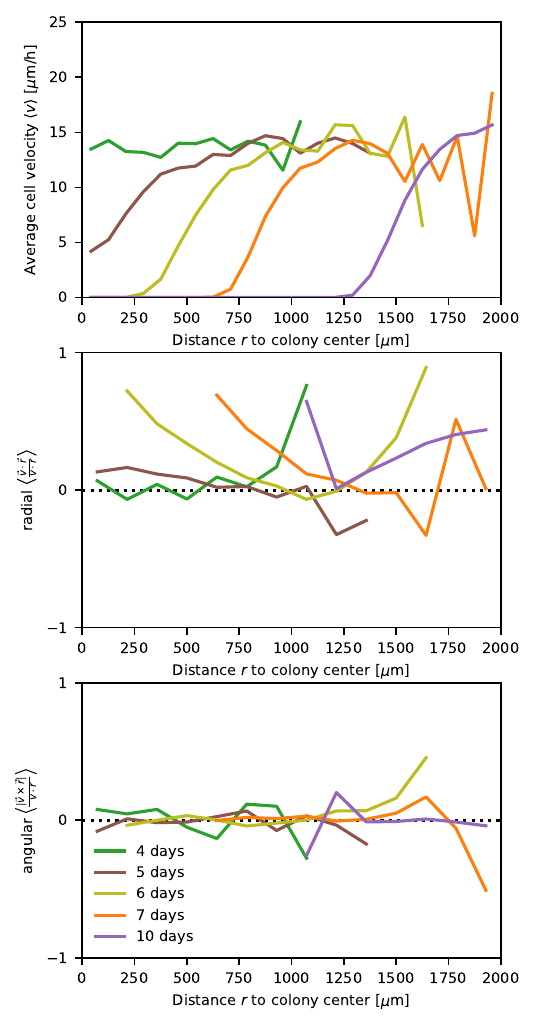}
    \caption{Average cell velocities in the calibrated CA with respect
      to distance $r$ from the center of the colony display radial
      velocity gradient (top panel) and radially outwards biased cell
      motion (middle and bottom panel). Shown are averaged the (top
      panel) absolute velocities $\left<v\right>$, (middle panel)
      relative radial velocity
      $\left<\frac{\vec{v} \cdot \vec{r}}{v \cdot r}\right>$, and
      (bottom panel) relative angular velocity
      $\left<\frac{|\vec{v}\times\vec{r}|}{v \cdot r}\right>$ at the
      time points $4, 5, 6, 7, 10$~d corresponding to
      \figref{app-direction} and \figref{direction}. While the angular
      velocity is always around zero (indicated by the dashed black
      line in the bottom panel), the radial velocity exhibits values
      well above zero.}
    \label{fig:radial-profile}
\end{figure}

\begin{figure}
    \includegraphics{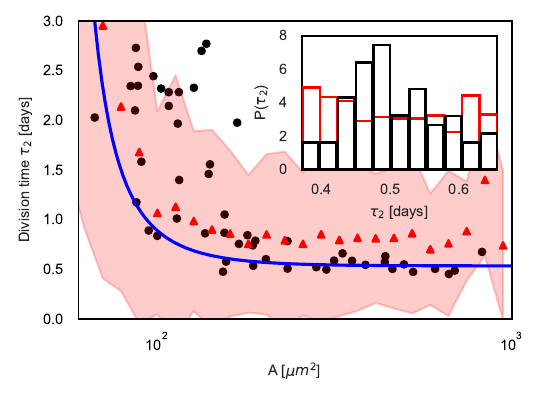}
    \caption{Emergent dynamics of the calibrated CA model (red
      triangles) reproduce the observed division times $\tau_2$ for an
      initially homogeneously seeded cells (black dots from
      Ref.~\cite[Fig.~4E]{PulHufNevStrSigFygShr2012}, blue line
      corresponds to the fitted proliferation rate from
      \figref{proliferation-rate}). The division times of the CA model
      are averaged over intervals of the cell size $A$ and
      corresponding standard deviation, resulting from the underlying
      exponential distribution of the division times, is shown as red
      shaded region. In the inset the according distributions of the
      division times of CA model (red) and experiment (black
      ~\cite[Fig.~4E inset]{PulHufNevStrSigFygShr2012}) are compared.
      While division times in the CA model follow an exponential
      distribution, biological cells divide at more regular intervals
      leading typically to a narrower distribution.}
    \label{fig:division-times}
\end{figure}

\begin{figure}
  \includegraphics{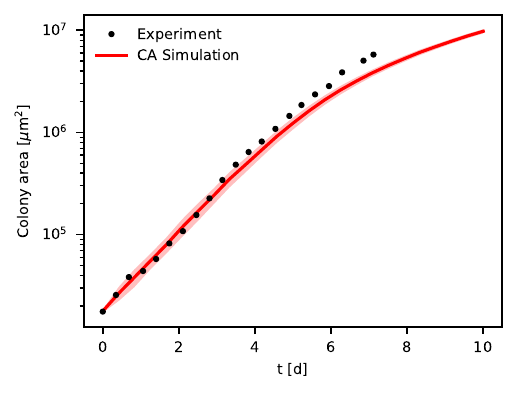}
  \caption{Emergent dynamics of calibrated CA model (red line) reproduces
    experimentally observed size of expanding epithelial colony (black
    dots from Ref.~\cite[Fig. 1B]{PulHufNevStrSigFygShr2012}). The
    colony size exhibits initially an exponential growth (low-density,
    confluent regime) and slows around time $t_T = 6 \pm 1$~d to a
    subexponential growth (transition to a stationary, postconfluent
    state). Results from CA model are averaged over $15$ independent
    runs and standard deviation is displayed as red shaded region.
    Model parameters as \figref{colony-area} in the main text, but for
    a smaller maximal migration velocity
    $v_\text{max} = 50 \ \mu\text{m} /\text{h}$.}
    \label{fig:app-colony-area}
\end{figure}

\begin{figure}
  \includegraphics[width=\linewidth]{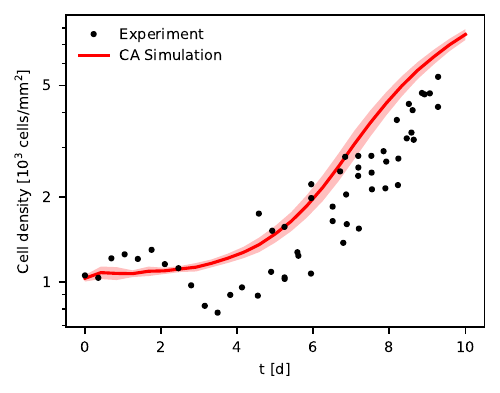}
  \caption{Emergent dynamics of calibrated CA model (red line) reproduces
    experimentally observed cell densities of expanding epithelial
    colony (black dots from Ref.~\cite[Fig.
    1C]{PulHufNevStrSigFygShr2012}, corresponding data to areas
    displayed in \figref{app-colony-area}). The cell density is
    initially constant (low-density, confluent regime) and increases
    rapidly around time $t_T = 6 \pm 1$~d (transition to a stationary,
    postconfluent state) implying dense packing of cells, which become
    smaller and smaller due to slowing but continued proliferation
    without cell growth. Results from CA model are averaged over $15$
    independent runs and standard deviation is displayed as red shaded
    region. Model parameters as \figref{colony-density} in the main
    text, but for a smaller maximal migration velocity
    $v_\text{max} = 50 \ \mu\text{m} /\text{h}$.}
    \label{fig:app-colony-density}
\end{figure}

\begin{figure}
  \includegraphics{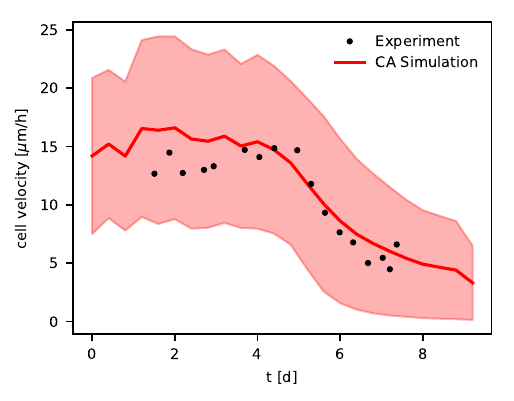}
  \caption{Emergent dynamics of calibrated CA model (red line)
    reproduces qualitatively the experimentally observed cell
    velocities of expanding epithelial colony (black dots from
    Ref.~\cite[Fig. 2C]{PulHufNevStrSigFygShr2012}, corresponding data
    to Figs.~\ref{fig:app-colony-area}
    and~\ref{fig:app-colony-density}). The average cell velocity is
    initially constant (low-density, confluent regime) and decreases
    rapidly around time $t_T = 6 \pm 1$~d (transition to a stationary,
    postconfluent state). Presented are the root-mean-square
    velocities, where for the model velocities are evaluated from a
    single simulation run over time increments of $580$~min to neglect
    small-scale oscillations and standard deviation is displayed as
    red shaded region. Model parameters as \figref{cell-velocity} in
    the main text, but for a smaller maximal migration velocity
    $v_\text{max} = 50 \ \mu\text{m} /\text{h}$.}
    \label{fig:app-cell-velocity}
\end{figure}

\begin{figure}
  \includegraphics{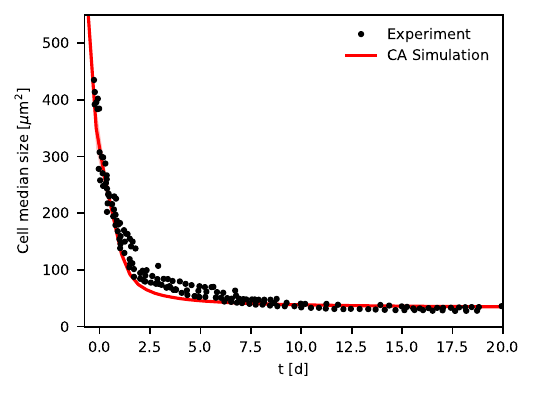}
  \caption{Emergent dynamics of calibrated CA model (red line)
    consistent with experimentally observed cell size median for
    initially homogeneously seeded cells (black dots from
    Ref.~\cite[Fig. 3C]{PulHufNevStrSigFygShr2012}, transition $t_T$
    to a stationary, postconfluent state corresponds roughly to
    $1.5$~d). Cell size median initially drops rapidly and then
    saturates at small cell sizes $A\sim 40 \ \mu\text{m}^2$. Model
    parameters as \figref{median} in the main text, but for a smaller
    maximal migration velocity
    $v_\text{max} = 50 \ \mu\text{m} /\text{h}$. Results from CA model
    are averaged over $15$ independent runs and standard deviation is
    displayed as red shaded region.}
    \label{fig:app-median}
\end{figure}
\end{document}